\pdfoutput=1

\documentclass[a4paper, reprint, amsmath]{revtex4-2}

\usepackage{graphicx}
\usepackage[caption=false]{subfig}
\usepackage{hyperref}

\begin{document}

\title{G-CoReCCD: a GPU-based simulator of the charge transport in fully-depleted CCDs}

\date{\today}

\author{Nicolas E. Avalos}
\author{Miguel \surname{Sofo Haro}}
\affiliation{CONICET, Centro Atómico Bariloche and Instituto Balseiro, Comisión Nacional de Energía Atómica (CNEA), Universidad Nacional de Cuyo (UNCUYO), Río Negro, Argentina.}

\pagestyle{plain} 

\begin{abstract}

We introduce a simulator of charge transport in fully-depleted, thick CCDs that include Coulomb repulsion between carriers. The calculation of this long-range interaction is highly intensive computationally, and only a few thousands of carriers can be simulated in reasonable times using regular CPUs. G-CoReCCD takes advantage of the high number of multiprocessors available in a graphical processing unit (GPU) to parallelize the operations and thus achieve a massive speedup. We can simulate the path inside the CCD bulk for up to hundreds of thousands of carriers in only a few hours using modern GPUs.

Published version available at \url{http://dx.doi.org/10.1117/12.2580606} \cite{avalos2020}.

\end{abstract}


\maketitle

\section{Introduction}

Scientific CCDs have been extensively used in ground and space-based astronomy and X-ray imaging. Furthermore thick fully-depleted CCDs fabricated on high resistivity silicon allow for increased detection mass, enabling their use as particle detectors in neutrino and dark matter experiments. In X-rays applications it is required to model the expected image that the interaction will produce to evaluate the resolution of the system. Also in particle experiments it is important to extract as much information as possible from background particles like muons, X-rays or electrons. The existing models used for imaging applications treat the charge transport in the depleted silicon only considering the carrier diffusion by the crystal thermal energy. With the previous simplification it is possible to give an analytical treatment to the problem. There is however experimental evidence that the Coulomb interaction between carriers that occurs during the collection process cannot be neglected \cite{haro2020studies}. In the case of X-rays there is an appreciable dependence of the events size in the image with the energy of the X-ray due to the Coulomb forces in the charge cloud produced by the X-ray. In precision astronomy, even though the amount of charge produced by each photon is small, the charge already stored in each pixel deflects the forthcoming charges affecting the acquired image \cite{guyonnet2015evidence}. Including the interaction between carries adds a complexity to the problem that does not allow for an analytical approach and a computational solution is necessary. In this work we present the first GPU-based simulator of the charge transport in a CCD. It was developed to simulate the evolution of up to millions of carriers and produce an output image starting from the initial distribution of the carriers. Some other works on simulated images can be found in \cite{lage2019,britt2022,lamarr2022}.

In Sec.~\ref{sec:transport} of this article we introduce the physical model for charge transport. In Sec.~\ref{sec:gpu} we briefly explain the basics of GPU programming and then the algorithm of the simulator. In Sec.~\ref{sec:performance} we show a comparison between the computation times in CPU and in GPU. Finally, in Sec.~\ref{sec:applications} we show applications for the simulator, which include X-rays and atmospheric muons interaction with the CCD.

\section{Charge transport in fully-depleted CCDs}
\label{sec:transport}

\begin{figure}[tb]
    \includegraphics[width=\linewidth]{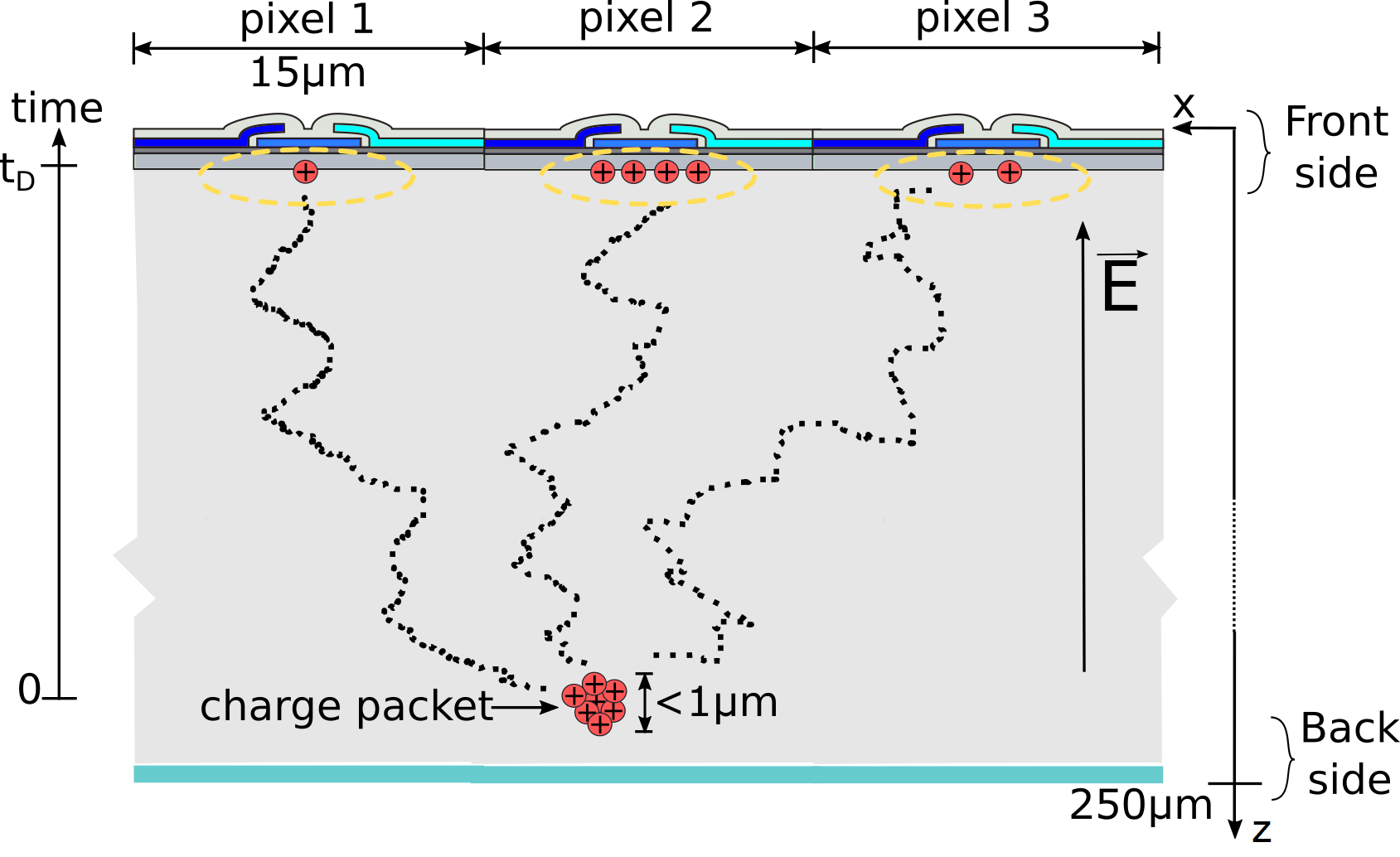}
    \caption{Depiction of the paths travelled by the charges in the CCD bulk until they arrive at the pixel potential well. Image was extracted from [\onlinecite{haro2020studies}].}
    \label{fig:charge_packet}
\end{figure}

Charge is generated in the CCD bulk whenever a particle deposits its energy and creates electron-hole pairs in the silicon. Photons, for example, interact via photoelectric effect, exciting the electrons in the valence band and sending them to the conduction band. Other type of particles can interact via nuclear or electron recoil, thus producing a similar effect. The created holes are afterwards pushed towards the CCD potential wells on the front side of the detector, as illustrated in Fig.~\ref{fig:charge_packet}. This is produced because of the depleting voltage, but there are also other mechanisms that produce charge transport, such as charge diffusion and Coulomb repulsion. We will now briefly discuss each of this mechanisms. A more complete discussion can be found at [\onlinecite{haro2020studies}].

\subsection{Drift due to depletion voltage}

The electric field $E$ in the bulk can be modeled as a linear function of the depth $z$:
\begin{equation}
    E(z) = a_1 z - a_2 \quad ,
    \label{eq:Estat}
\end{equation}
with $a_1$ and $a_2$ constants given by the construction characteristics of each specific CCD such as dopant concentration in the silicon bulk, thickness of the p-channel and the field in the p-n junction. The holes will then drift towards the potential well with a velocity following the rule $\vec{v} = \mu \vec{E}$. The proportionality constant $\mu$ is the hole mobility, which depends on both the absolute value of the electric field $E$ and the temperature $T$ of the CCD:
\begin{equation}
    \mu(E,T) = \frac{1.31\times10^8 T^{-2.2}}{\left[ 1 + \left( \frac{E}{1.24T^{1.68}} \right)^{0.46T^{0.17}} \right]^{1/(0.46T^{0.17})}} \quad ,
    \label{eq:mu}
\end{equation}

\subsection{Charge diffusion}

Since the CCD absolute temperature is not zero, the silicon atoms have thermal kinetic energy. The charge carriers scatter with these atoms in their path to the potential well. The average free path between collisions is small compared to the full pixel depth, so the movement can be characterized as a Brownian motion process. Therefore, at each time interval $t_D$ there will be a random displacement vector $\Delta \vec{r_D}$ whose coordinates will be given by a normal distribution with mean zero and variance $\sigma_D^2$, calculated via the Einstein equation of difussion:
\begin{equation}
    \sigma_D^2 = 2 D_h t_D \quad ,
\end{equation}
where $D_h$ is the holes diffusion coefficient:
\begin{equation}
    D_h = \frac{k_B \mu_h T}{q} \quad ,
\end{equation}
with $k_B$ the Boltzmann constant and $q$ the electron charge.

\subsection{Coulomb repulsion}

This effect is produced whenever two carriers with charge of the same sign are close. In this case, we have many carriers with the same charge drifting towards the potential well. Each of these generates an electric field and is affected by the sum of the field generated by every other charge.

Each carrier $i$ produces an electric field $\vec{E}_i$. Assuming we have $N$ carriers, the electric field produced by Coulomb repulsion at the position $\vec{r}_j$ (corresponding to the $j$th carrier) is:
\begin{equation}
    \vec{E}_{Coul} (\vec{r}_j) = \sum_{i \neq j} \vec{E}_i(\vec{r}_j) = \frac{q}{4 \pi \epsilon_{Si}} \sum_{i \neq j} \frac{\vec{r}_j - \vec{r}_i}{|\vec{r}_j - \vec{r}_i |^3} \quad .
    \label{eq:Edin}
\end{equation}
This effect will be larger as the number of charges is increased and the closer they are from one another.

\section{GPU implementation}
\label{sec:gpu}

\subsection{GPU programming basics}

The main advantage of programming applications in GPUs is a massive improvement in the time performance of an algorithm. This happens because the GPUs have multiple cores that can perform similar computations simultaneously. This parallelism, however, is detrimental to the communication between processes, making it hard to exchange information between them. Therefore, it is more common to use heterogeneous computing in which serial and parallel processes happen sequentially. A very powerful tool that is useful to program an heterogeneous application in GPU is CUDA \cite{cuda}. With this platform, applications can be programmed using common programming languages such as C/C++, FORTRAN or Python.

Within this framework, a function that is executed in the GPU is called a kernel. Whenever a kernel is called, one has to define a grid with blocks and a number of threads per block. One can also assign a certain amount of shared memory per block. Communication between CPU (host) memory and GPU (device) memory is very slow, and thus copies across them should be kept as scarce as possible. The GPU has several types of memories. The biggest is the global memory, an this can be accessed from any thread at any given block. There is also a per-block shared memory, which is only accessible by threads in the corresponding block, and reading and writing processes are faster than those produced in the global memory.

\subsection{The algorithm}

\begin{figure}[tb]
    \includegraphics[width=.7\linewidth]{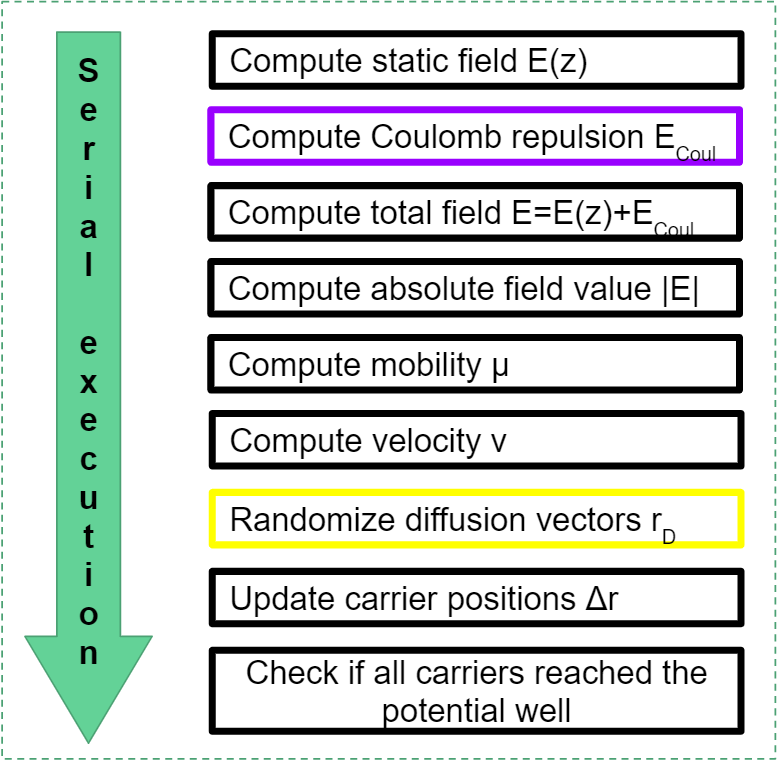}
    \caption{Sequence of computations at the core of the simulator. Each rectangle represents a parallel computation, and the up-to-down arrangement depicts the sequential order in which they are realised. These are iterated in loop for each time step $dt$, and stop when all the charges have arrived at the potential well. In black rectangles are shown the computations programmed using Thrust library, in yellow the ones using cuRAND library and in purple the ones coded using a custom CUDA kernel. }
    \label{fig:algorithm}
\end{figure}

Fig.~\ref{fig:algorithm} depicts the computations executed in the main algorithm of the simulator. Each rectangle represents a computation performed in the GPU in a parallel way. In the vertical direction, the sequential order of the latter computations is shown. Assume we start at time $t=0$ and we have a time step $dt$. We already have loaded to the GPU the position vectors of each charge. The operations described below are computed parallelly for each of the charges simulated. First of all, the algorithm computes the electric field given by depletion $E(z)$ (eq.~\ref{eq:Estat}); afterwards, the field given by Coulomb repulsion $\vec{E}_{Coul}$ (eq.~\ref{eq:Edin}) and then the sum of both. With this sum, it computes its absolute value $|E|$, the mobility $\mu(E,T)$ (eq.~\ref{eq:mu}) and the velocity $\vec{v}$. It then initialises the random diffusion vectors $\Delta \vec{r}_D$ and computes the total displacement $\Delta \vec{r}$. Finally, it checks the $z$ positions of every charge. If all charges have arrived at the potential well then the loop exits; else, it increases the time $t = t + dt$ and repeats the loop. 

The colors in Fig.~\ref{fig:algorithm} depict the tools used to program each process. The code makes extensive use of the Thrust parallel library \cite{thrust}, shown in black in the figure. This library works similarly to the C++ standard library, but is optimized for GPU computing. In yellow is shown the operation that uses the cuRAND library \cite{curand}, that is, the initialization of random vectors. Finally, in purple is shown the operation that was programmed using a custom CUDA kernel, the calculation of the Coulomb repulsion field, and we will explain this one in detail below.

\begin{figure}[tb]
    \includegraphics[width=\linewidth]{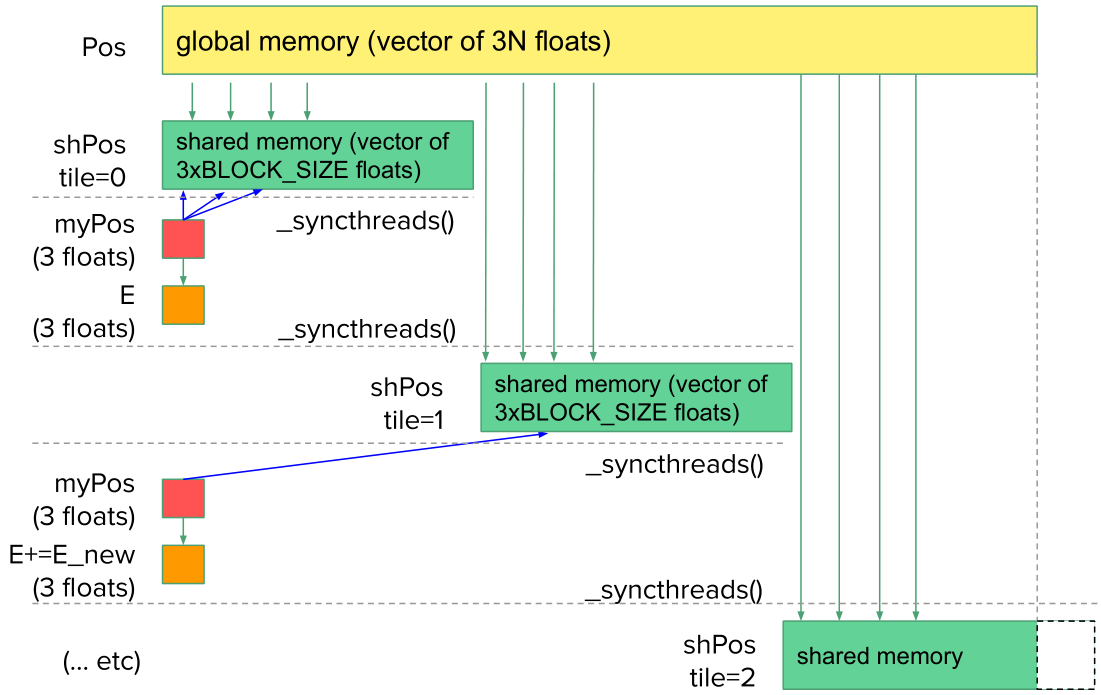}
    \caption{Schematic of the Coulomb repulsion calculation within the kernel. \texttt{Pos} is the vector of $3N$ floats that contains the 3D positions of the $N$ carriers simulated. Each thread in a block is assigned a carrier and loads its position vector \texttt{myPos} to the shared memory vector \texttt{shPos}. The GPU then ensures that all threads have finished their work (\texttt{\_syncthreads()} built-in function) before starting with the calculation of the Coulomb field for each charge. The resultant vector \texttt{E} is saved to global memory. Again the \texttt{\_syncthreads()} function is called and afterwards the threads load a new carrier position to shared memory and the process is repeated. The iteration continues until all the charges have been loaded to shared memory and their contribution to the field \texttt{E} computed. }
    \label{fig:kernel}
\end{figure}

The Coulomb charge repulsion effect is long-range and thus the computational complexity for its calculation is $\mathcal{O}(N^2)$. Here, each GPU thread is assigned a charge position and then computes the distance to every other charge in the simulation. A softening factor $\epsilon^2$ is added at this point in order to avoid diverging results produced by charges standing too close to each other and take the self-interaction to zero:
\begin{equation}
    \vec{E}_{j}^{Coul} = K \sum_{i=1}^N \frac{\vec{r}_{ij}}{(|r_{ij}|^2 + \epsilon^2)^{3/2}} \quad ,
    \label{eq:softening}
\end{equation}
where $K=q/(4 \pi \epsilon_0 \epsilon_{Si})$ and $\vec{r}_{ij} = \vec{r}_j - \vec{r}_i$, being $\vec{r}$ the position vector of each charge. In order to increase the speedup and avoid multiple reading and writing to the GPU global memory, we implemented the tile calculation described in Chapter 31 of \cite{gg3-nbody}, where the position vectors are divided into smaller pieces called \textit{computational tiles}. Fig~\ref{fig:kernel} shows a schematic of the actions performed by the kernel. First, each thread in a particular block loads the position vector of a carrier from global memory and saves it to shared memory. Once all threads finish this task, the calculation described in eq.~\ref{eq:softening} is performed and the resulting field vector $\vec{E}_{Coul}$ is saved to global memory. However, since only a limited number of charge positions were loaded to shared memory (because of the limitations in the allowed amount of threads per block), not all the carriers were taken into account yet. Thus the algorithm advances to the next \textit{tile}, were new charge positions are loaded to shared memory. The resulting field is added to the previously saved. This sequence is continued until all the charges are processed.

The simulator allows for user input to customize the simulation. When the program is executed with no arguments, random initial position vectors are initialized, simulating an X-ray interaction, and prompts for inputs appear in the command line. The required input includes the number of carriers $N$ to simulate and the initial depth of the interaction, the CCD properties including its thickness, the electric field parameters $a_1$ and $a_2$ from eq.~\ref{eq:Estat} and the depth of the potential well. It also prompts for the maximum number of iterations desired, the time step $dt$, whether Coulomb repulsion should be computed and asks whether final results should be saved to a file. The program can also be called with arguments, in which case one can load the charge positions and/or the simulation parameters from an external file.

\section{Performance results}
\label{sec:performance}

\begin{figure*}[tb]
    \subfloat[Coulomb repulsion OFF]{
        \includegraphics[width=0.37\linewidth]{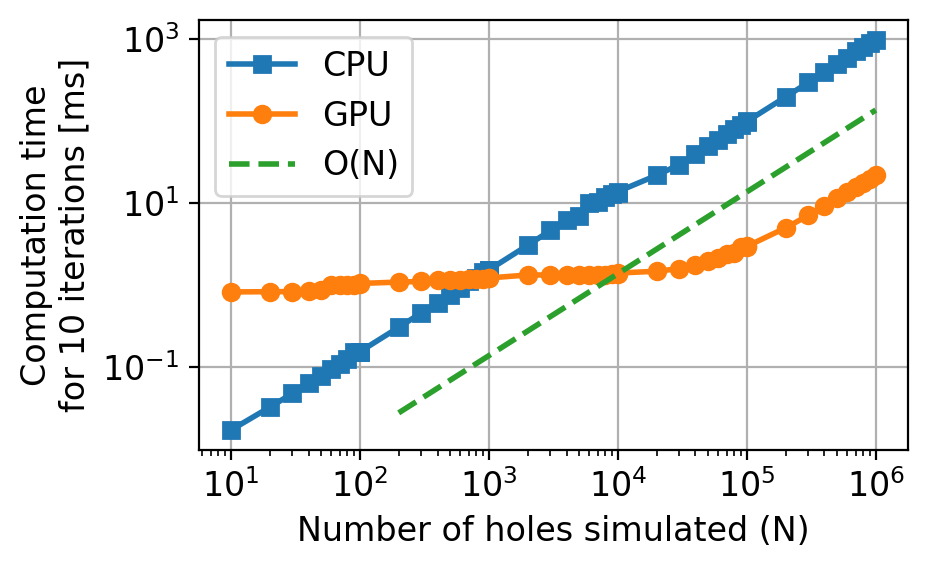}
        \includegraphics[width=.37\linewidth]{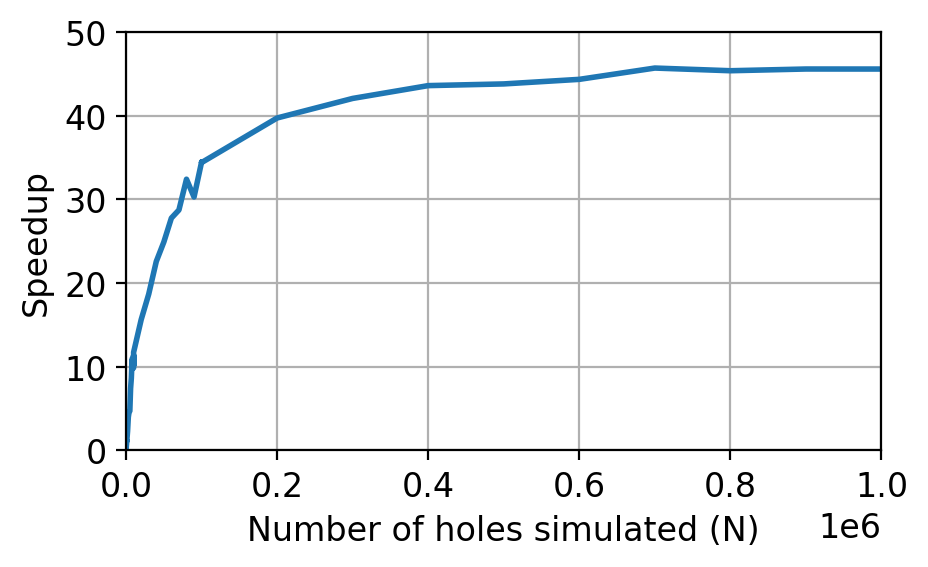}}
    \\
    \subfloat[Coulomb repulsion ON]{
        \includegraphics[width=0.37\linewidth]{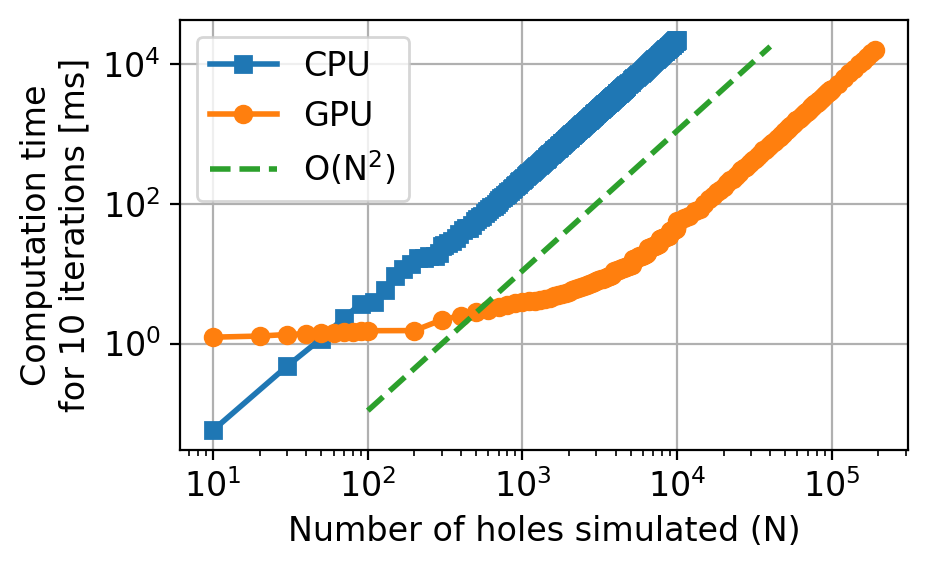}
        \includegraphics[width=.37\linewidth]{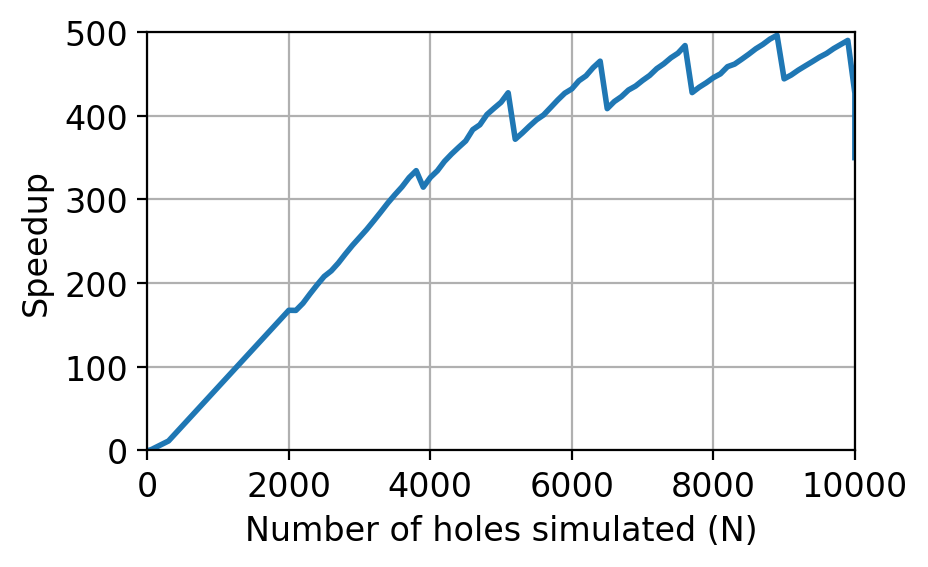}}
    \caption{\textbf{(Left)} Comparison between computation time for 10 iterations of the simulator algorithm in an AMD Ryzen 7 3700X 8-Core Processor CPU (blue squares) and a NVIDIA Geforce RTX 2070 Super GPU (orange circles). In dashed line we show the expected behavior for $\mathcal{O}(N)$ and $\mathcal{O}(N^2)$ processes. \textbf{(Right)} Speedup of the algorithm when running on GPU. }
    \label{fig:xpuPerf}
\end{figure*}

We show in the left side of Fig.~\ref{fig:xpuPerf} a comparison of the computation time when running 10 iterations of the simulator algorithm in an AMD Ryzen 7 3700X 8-Core Processor CPU and in a NVIDIA Geforce RTX 2070 Super GPU. The same test was run disabling (top graphs) and enabling (bottom graphs) the Coulomb interaction between charges. In dashed line we show a reference for the expected behavior of each computation performance, $\mathcal{O}(N)$ in the first case and $\mathcal{O}(N^2)$ in the second. We observe that the time for the CPU has the same behavior in the whole range of the number of holes simulated, but the time for the GPU has two different behaviors. For a low enough number of charges, the computing time is close to constant. This means that the GPU is sub-utilised and there are idle processors. If we simulate a high enough number of charges, however, the behavior is the same as the CPU. We can see that the GPU performs better for simulations of approximately 60 charges or more if Coulomb repulsion is computed, whereas if it is not the performance is better when computing at least 1000 charges.

On the right side of Fig.~\ref{fig:xpuPerf} we show the speedup, defined as the time the computation took in the CPU divided by the time it took in the GPU. We can see that it reaches an asymptotic value of 45X when not computing Coulomb repulsion. The speedup is greatly higher when we do compute Coulomb interactions, reaching up to 500X. The saw-teeth shape of the curve in this case is related to the occupation of the GPU, which varies because some threads may be idle during the computation if the number of charges simulated is not an exact multiple of the number of threads used. The exact shape and value of this curves are highly dependent on the model of the CPU and GPU used, and may too depend on code optimizations and the compiler used.

\section{Applications}
\label{sec:applications}

\subsection{X-rays}

X-rays interact via photoelectric effect with the silicon and produce a number of charges determined by its energy $E_X$. The energy of the photo-electron produced is $E_{e^-}=E_X-E_B$, with $E_B$ the binding energy. The number of charge carriers produced ($N$) follows the rule $N=E_{e^-}/w_{e^-}$, where $w_{e^-}$ is the average energy required to generate an electron-hole pair in silicon. This average energy depends on the temperature of the CCD and is equal to $3.77$\,eV at 140\,K.

The initial cloud size of charge carriers can be modeled following a 3D Gaussian distribution with standard deviation depending solely on $E_{e^-}$ (a derivation of this model can be found elsewhere \cite{yousef2011energy,williams2001x,ashley1976calculations}). Thus, we can simulate the initial positions of the carriers of an X-ray of specific energy, simulate its drift across the CCD bulk and find its final spread. Then we can compare these results with experiments, and a good agreement with the simulation has been found in a previous work \cite{haro2020studies}.

\begin{figure*}[tb]
    \includegraphics[width=.37\linewidth]{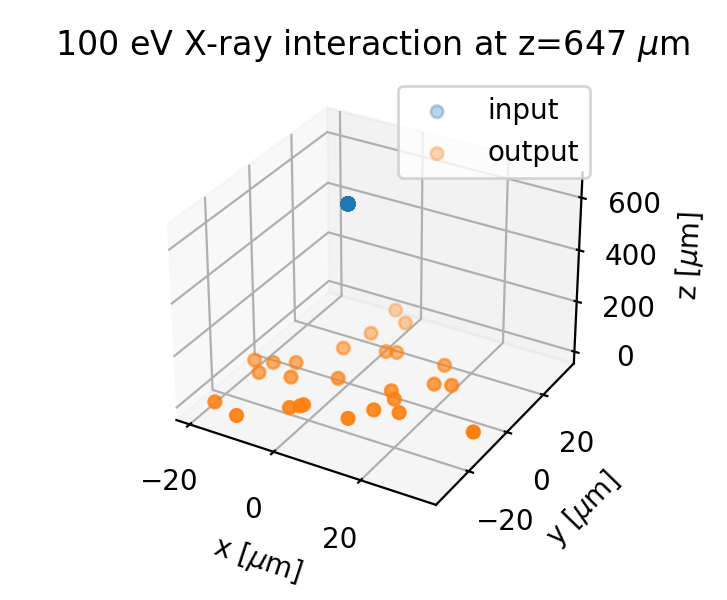}
    \includegraphics[width=.37\linewidth]{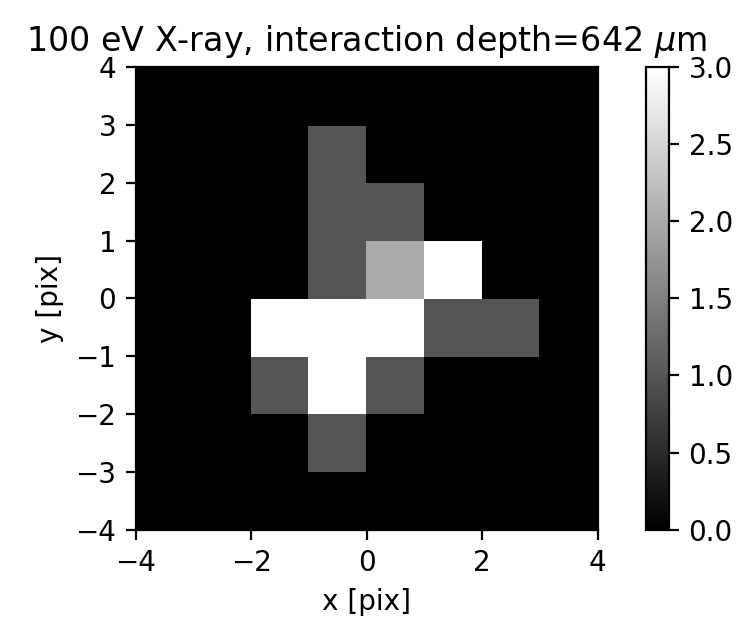}
    \caption{\textbf{(Left)} Starting positions of the charges produced after an interaction of a 100\,eV X-ray with the CCD at 647\,$\mu$m deep (blue dots) and final positions after the G-CoReCCD simulation (orange dots). \textbf{(Right)} Simulated image obtained by computing a 2D histogram of the output positions, with bin size equal to the pixel with (15\,$\mu$m in this case).}
    \label{fig:xrays}
\end{figure*}

On the left side of Fig.~\ref{fig:xrays} we show a 3D plot of the initial and final positions of carriers for a 100\,eV X-ray interaction in a CCD at a depth of 647\,$\mu$m. On the right side of the figure we show a simulation of the image produced by such X-ray, considering a pixel width of 15\,$\mu$m. We can see that, although the event was initially point-like, the spread of the charges became much bigger when they arrived at the potential well.

\subsection{Atmospheric muons}

\begin{figure*}
    \includegraphics[width=.3\linewidth]{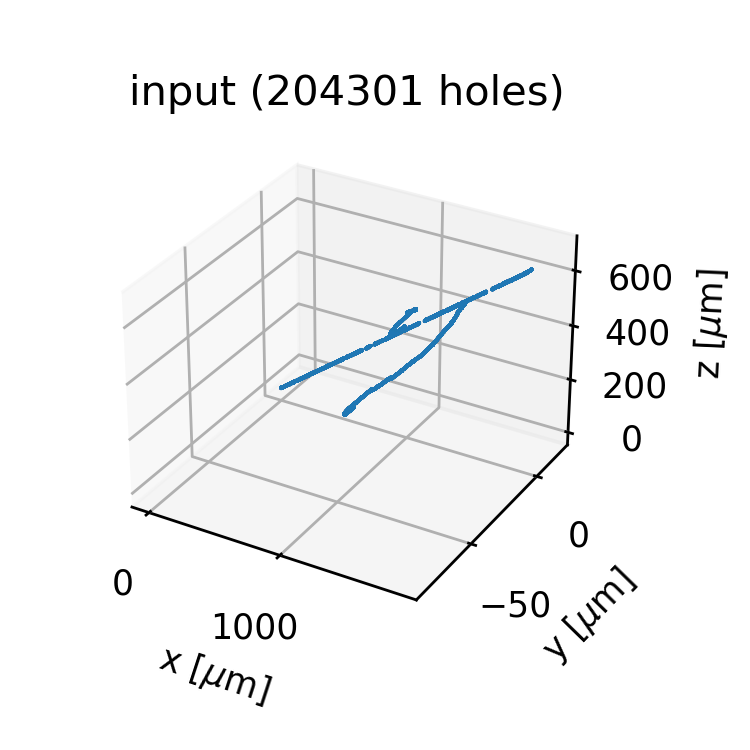} 
    \includegraphics[width=.63\linewidth]{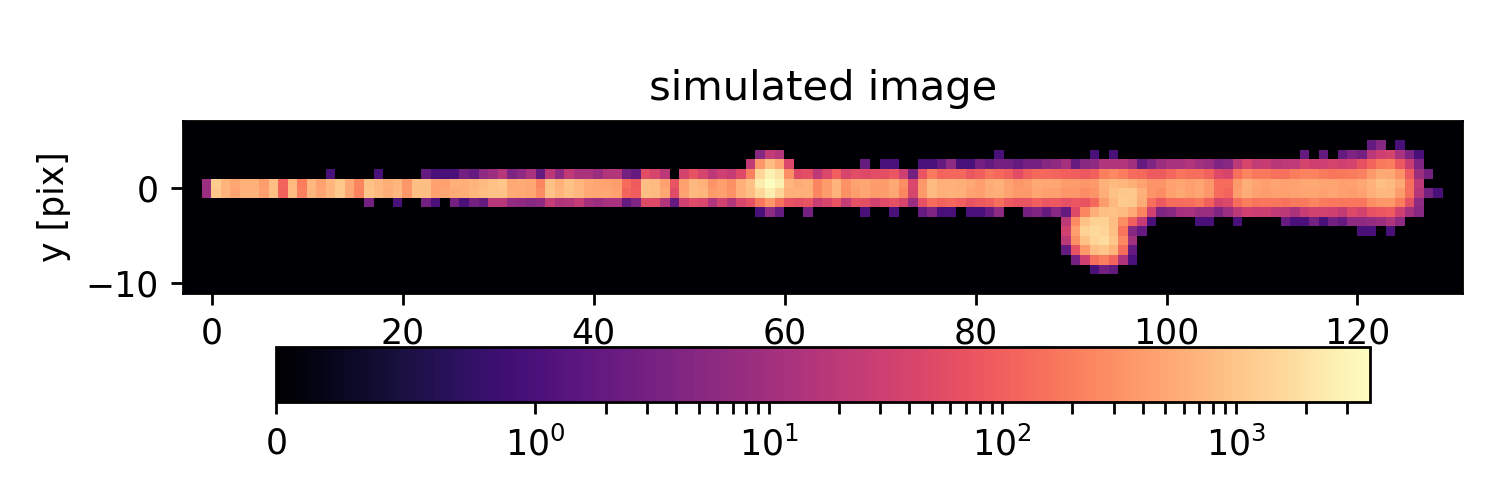}
    \caption{\textbf{(Left)} Initial distribution of holes in the CCD after an interaction with a 3\,GeV atmospheric muon. See text for details. \textbf{(Right)} Simulated image of the muon interaction obtained with G-CoReCCD. }
    \label{fig:muons}
\end{figure*}

Produced when high-energy cosmic rays interact with the atmosphere, muons account for big sources of background when carrying a particle detection experiment at surface level. This charged particles leave a characteristic straight track across the CCD, because they are so energetic that very little energy is lost during the interaction. In fact, they may cause ionization in the silicon, generating free electrons in the bulk as they pass. Fig.~\ref{fig:muons} shows to the left a simulated track of a 3\,GeV muon inciding at 70$^\circ$ across a 675 $\mu$m CCD, obtained with the software GEANT4 \cite{geant4}. Secondary electrons are produced because of ionization and appear as worm-like tracks. The total number of electron-hole pairs produced was 204301, and their initial positions were fed to the G-CoReCCD simulator. The right side of Fig.~\ref{fig:muons} shows the image that would be produced at the CCD. It replicates well the low spread of the charges produced near the front end and the bigger spread of the charges produced near the back end, which had to travel all the way down to the pixel potential well. The colorscale is logarithmic, and shows how many charges are found in each pixel. In a NVIDIA GTX R2070 SUPER GPU, the full computation of the transport of this charges takes about 4~hours. The same computation carried in a CPU would be nearly impossible, as extrapolating the results of Sec.~\ref{sec:performance} we obtain that it would take about 83 days to complete.

\section{Conclusions} 
\label{sec:conclu}

We have developed a simulator for charge transport in fully-depleted, thick CCDs that takes into account the Coulomb repulsion between charges and performs computations parallelly in GPU devices. A significant acceleration with respect to a similar code that runs in CPUs was found, obtaining a 45X speedup for processes of complexity $\mathcal{O}(N)$ and up to 500X speedup for $\mathcal{O}(N^2)$ complex processes. We have tested the code for applications in X-rays and atmospheric muons, obtaining simulated images that correctly replicate the expected particle tracks left in the CCDs.

\section*{Acknowledgments}

We would like to thank Guillermo Fernández Moroni (UNS-CONICET) for providing the CPU version of the software from which our GPU code was derived, Fabricio Alcalde Bessia (CNEA-CONICET) for providing the GEANT4 muon simulation and Xavier Bertou (CNEA-CONICET) for his permanent support.

\bibliographystyle{spiebib2}
\bibliography{
bibdoi
}

\end{document}